# Studying the Affecting Factors on Trust in Social Commerce

Leila Esmaeili

Ph.D. Candidate, Computer Engineering and Information
Technology Department
Amirkabir University of Technology
Tehran, Iran

Shahla Mardani

M.Sc., Computer Engineering and Information Technology
Department
Amirkabir University of Technology
Tehran, Iran

Muhamad Mutallebi

M.A. Student, Entrepreneurship Department
University of Tehran
Tehran, Iran

Seyyed Alireza Hashemi Golpayegani

Assistant Professor, Computer Engineering and Information
Technology Department
Amirkabir University of Technology
Tehran, Iran

*Abstract*—**Recently, e-commerce has enjoyed the appearance of many novel opportunities created as a result of the increasing growth of social networks. Social commerce (s-commerce), an offspring of e-commerce, interconnects users and helps with the commerce process facilitated by social networks as well as other media. Uncertainty and trust are always an issue in any type of commerce and especially in s-commerce where, due to its nature, a higher level of risk and uncertainty exists. Despite the history for trust in virtual commercial transactions, there are a limited number of researches on the topic; and the current shift towards commerce in virtual communities and networks begs further attention. Therefore, in this paper, by scrutinizing the available literature on the topic, the affecting factors on trust in s-commerce are identified and introduced. Among the reviewed papers, 7 papers matched the criteria, and overall, 19 factors were extracted. Firm size, trust disposition, information quality, and familiarity with other members were among the most cited factors in the literature.**

## I. INTRODUCTION

The emergence of web 2.0 technologies and its wide acceptance in social network service market, facilitated online shopping by means of some very special and interactive methods; this caused e-commerce to gradually shift to s-commerce [1]. S-commerce refers to the use of web 2.0 technologies in traditional e-commerce [2]; the features and impacts of web 2.0 on e-commerce are discussed in the literature such as in [3] and [4]. Empirical evidence shows the use of a set of features by web-based businesses to increase the social interaction and sense of cooperation among users in the late 90s [1][5]; however, an official use of the term "social commerce" has been introduced to the literature by the first use of the term by Yahoo in 2005 [6][5][7] and its premier appearance in a scientific paper goes only as far back as 2007

[8][9]. S-commerce has ever since drawn a considerable attention [5][7] and has been known rather by its practical applications and not academically [10].

Commerce and commercial transactions have essential elements to them regardless of the type of commerce (electronic or not and social or not); one of these elements is trust without which conducting financial transactions proves problematic (see section II). Trust is the willingness of one party to put himself in a position of harm by another party, expecting that the other party will do actions that are important to him, even in the absence of control and monitoring [11]. In general, trust is considered an essential factor in all financial and social circumstances; but especially in commerce, due to its impact on consumers' behavior and its role in strengthening their purchase intention, the importance of trust is regarded even more vital [12]. Furthermore, the vital importance of trust is further stressed in e-commerce when the risk of transactions is higher than normal situations and customers face an uncertain environment [13] (see section III).

To reduce uncertainties in e-commerce and help improve trust, researchers have recommended relationships among people and knowledge transfer as an effective solution [14][15][16]. Knowledge transfer is now made possible in social media platforms; and as previously discussed, combined with e-commerce social media shape the concept of s-commerce. This paper seeks to illustrate what factors influence trust in social commerce platforms and what is the nature of their influence?

In the following, the concept and definitions for s-commerce and trust, and also a review of literature on trust in social commerce is provided in section II. Section III introduces the research methodology and the affecting factors on trust in social commerce are discussed in section IV.

Finally, section V concludes the paper and provides avenue for future research.

## II. BASIC CONCEPTS AND DEFINITION

It is imperative to build a common definition for trust and what it means in this paper. Trust is constantly apparent in people lives and it shapes their behavior. The broad nature of the concept of trust makes it challenging to give a simple explicit description. A literature review of trust, social commerce, and trust in s-commerce is further explored in the following.

### A. Trust

Trust is a complex and multifaceted concept and holds many definitions [17]; nonetheless, there are some general definitions [18]:

*Definition I-reliability trust*: trust is a subjective probability and based on it, one party expects another peer to do an action in a way that satisfies the first party, the trusted.

Here, the trusted party is central to the definition and the trust of trusting party depends on reliability of the trusted party. However, in real life, trust does not exclusively depend upon the reliability of the trusted party. Falcone & Castel franchi discuss the same topic in [19] and do not consider high reliability as a sufficient factor for decision making. Therefore, another definition for trust is suggested.

*Definition II-decision trust* is the willingness of one party to trust another person or thing in a particular situation with a relative sense of security and awareness of the probability of negative consequences.

In this definition, trust and reliability as well as results and benefits are relative. In fact, here an attitude of risk is introduced. This definition gives a more comprehensive definition compared with definition I. Its implied uncertainty considers all possible consequences.

Complex as it is, trust possesses some generic characteristics [20][21]. To name some of these characteristics, trust is: directed, subjective, context specific, measurable, dependent on history, multi-faceted, dynamic, conditionally transferable, constitutional, and asymmetric. Further details of these characteristics are explained by in [22]

### B. Social Commerce

Although there is a lack of standard definition for s-commerce, it generally refers to providing and carrying out e-commerce activities and transactions in the context of social media and social networks in particular; therefore, by definition, s-commerce could be considered as a subset of e-commerce [23] or as its evolved form [24][5]; and it mainly includes social activities and commerce [23]. Marsden [25] has collected 22 different definitions of social commerce. They include a range of s-commerce features such as word of mouth advertising, trusted consultation, and assisted shopping. S-commerce is defined in respect to different fields of study like marketing, computer science, sociology, and psychology. In marketing, for instance, s-commerce regards the prominent trends in online markets; businesses use social media or web 2.0 as a direct marketing method in order to support customers' decision making process and behavior [26]. Regarding computer science, Lee et al. [27] Introduce s-commerce as an online application that combine web 2.0 technologies like Ajax [28] and RSS [29] with interactive platforms such as online social media websites and content communities in a commercial environment. Regarding sociology, s-commerce is about employment of web based social communities by e-commerce companies. It mainly focuses on the effects of social influence which shapes the interaction among consumers [2]. Finally, Marsden [30] from a perspective of psychology defines s-commerce as social shopping. In social shopping people intend to shop online under the influence of the prominent information in a social network. Based on the mentioned as well as other definitions, in this paper we define s-commerce as an Internet based commercial application that makes use of web 2.0 technologies and social media; and it supports user created content and social interactions. S-commerce considers a network of buyers and sellers as a single platform that includes the buying/selling activities and all related interactions. All of these commercial activities and transactions are carried out inside virtual communities and online markets.

### C. Trust in Social Commerce

The nature of e-commerce refrains sellers and customers to have a direct face to face transaction, and customers are not able to touch and feel the products. Face to face relations give both sides of the deal an opportunity to evaluate and learn about each other. Online transactions lack such opportunity and financial transactions usually carry a sense of uncertainty and risk; this is more of a problem when there has been no prior history for transaction between the peers. Social exchange theory states that people participate in transactions based on trust [31]. Therefore, as offline commerce requires trust, in online businesses, likewise, trust is an essential ingredient to reduce uncertainty.

As discussed, s-commerce is a novel phenomenon and research topics of a bibliography study of 2004 to 2013 mainly cover: study of customer/consumers behavior [32][33][34], s-commerce acceptance [35][36], and design of s-commerce websites [1][3]. In the first two topics, the issue of trust is a main concern of research. Over all, researches on this topic cover two subjects: (1) identification of effective factors on consumers' trust, and (2) studying the effects of various features of s-commerce on consumers' trust. In addition, another stream of researchers have studies the methods for improving consumers' trust on an s-commerce platform like word of mouth marketing and user experience.

## III. METHODOLOGY

In this section, the previous literature on s-commerce about trust is briefly reviewed based on data collection and analysis methods [37].

For the purpose of our work, we looked for researches in years between 2004 and 2013 on e-commerce and we particularly reviewed papers on the topic of trust.

To extract papers on trust in s-commerce, Google Scholar search engine was employed. Google Scholar, along with its simple search tools, provides researchers with more advanced search options of looking into papers, theses, books, and publications. Some main resources include IEEE, ACM, Springer, and Science Direct [38]. We used "s-commerce" and "trust" keywords in 2004 to 2013 in the publications field. The results were further refined to exclusively list full-text publications in English. Since there was a limited number of results (5 journal papers), papers on trust in commercial social networks or those focusing on the effects of social networks on trust in e-commerce were also added to the list. A total of 7 papers were selected and reviewed in our survey study.

Regarding [23], [39], and [40], table I gives an overview of all the papers used in study along with their main topics, research methodology, modeling method and analysis, data set, and data collection techniques.

Researches on trust in s-commerce have been carried on after 2009 and most of the works have been done in 2013; this illustrates the newness of the topic and the increasing interests of the academia on s-commerce. In most cases research models are based on the literature and data set required for analyzing the model have been obtained through (online or offline) questionnaires from a limited number of subjects. As a results, due to the sampling issues in some researches, the results may not be generalizable. Furthermore, in most studies, the survey methodology has been employed [33][41][42][43] and only in few other papers have researchers used other methodologies such as focus group and experimental research [44][45]. The researchers in their papers have used different theories like theory of planned behavior (TPB), technology acceptance model (TAM), trust transference theory, signaling theory, and the warranting principle to relate variables such as disposition to trust, user review, and subjective norm in their models to other factors and trust.

## IV. AFFECTING FACTORS ON TRUST

Building on each paper in section 3.1, figure II briefly shows which factors affect trust in s-commerce based on the analyses and the models provided in these papers. 18 factors were extracted from 7 different research papers, all of which have proved to have effects on trust in s-commerce. Out of 18 extracted factors five factors were cited in more than one paper. Except for one factor (negative user review), all factors had a positive causal relationship with trust in s-commerce.

## V. CONCLUSION

S-commerce is an evolved form of e-commerce [5][10][24], and it connects users through social networks and other social media, and facilitates the commerce process [42]. Many scientists consider s-commerce the center of the next wave of e-commerce [7][23][46]. Trust, as a result of the existing uncertainty and risk, is more obviously required when doing business on e-commerce platforms. Therefore, researchers have entered the subject and looked for methods to reduce this uncertainty by improving trust.

In the study of trust, decisions are made based on sharing of information about a certain peer by other people. Information sharing, user created content, and interacting with other peers, are the three very important features in social networks; and since s-commerce is a combination of web 2.0 technologies (and especially social networks) and commercial activities in a social and collaborative platform, it has created an opportunity to study trust and to improve it. Doing commercial activities in a social platform provides the stakeholders (customer, seller, firms, and intermediaries) with the information needed to evaluate the reputation and trust of users.

In this paper, after introducing the basic concepts of trust, previous researches on trust in s-commerce were studied; the results show that trust propensity, size of firm, information quality, familiarity with other members, and word of mouth referral are the most cited factors in the literature. Since there are a diverse range of factors influencing trust, and because trust also does influence other important factors that affect consumers' decision making process in s-commerce, more systematic researches are required.

For future work, it is recommended to investigate the proposed affecting factors on trust in s-commerce in an Iranian population with diverse demographic and geographic characteristics.

TABLE I.        CLASSIFICATION OF PAPERS STUDYING AFFECTING FACTORS ON TRUST IN S-COMMERCE

| ID | Author; year | Research theme | Research method | Modeling/analyzing method | Dataset; size | Data collection method |
|---|---|---|---|---|---|---|
| 1 | Gan et al.; 2009 [47] | Graphical modeling of trust | N/a | Graphical representation | N/a | N/a |
| 2 | Utz et al.; 2012 [44] | Effects of online reviews on trust | Experimental research | Regression analysis | University of Amsterdam students; 100 students for the first experiment and 131 users for the second experiment | Manual; no detail |
| 3 | Kim and Noh; 2012 [41] | Influencing factors on consumers' trust and trust performances | Survey | Research model, SEM | Korean s-commerce users; 466 users | Manual; online, offline, telephone and email methods |
| 4 | Leeraphong and Marajo; 2013 [45] | Trust and risk in purchase intention | Focus group | Research model | Working adults with 25 to 34 years of age; 15 users | Manual; no detail |
| 5 | Kim and park; 2013 [42] | Effects of s-commerce factors on trust and trust performance | Survey | Research model, PLS, SEM | Users living in Korea; 371 users | Manual; online and offline questionnaires, email, telephones, and random distribution of 2000 questionnaires among Korean users |
| 6 | Hajli et al.; 2013 [43] | Word of mouth and trust | Survey | Research model, SEM | Forum users, virtual communities, and social networks; 295 users | Manual; sending invitations through email to users in various networks in 2 months |
| 7 | Ng; 2013 [33] | Culture and intention to purchase | Survey | Research model, covariance-based structural equation modeling (CBSEM) | Facebook social network; 248 users | Manual; Sampling using snowball technique |

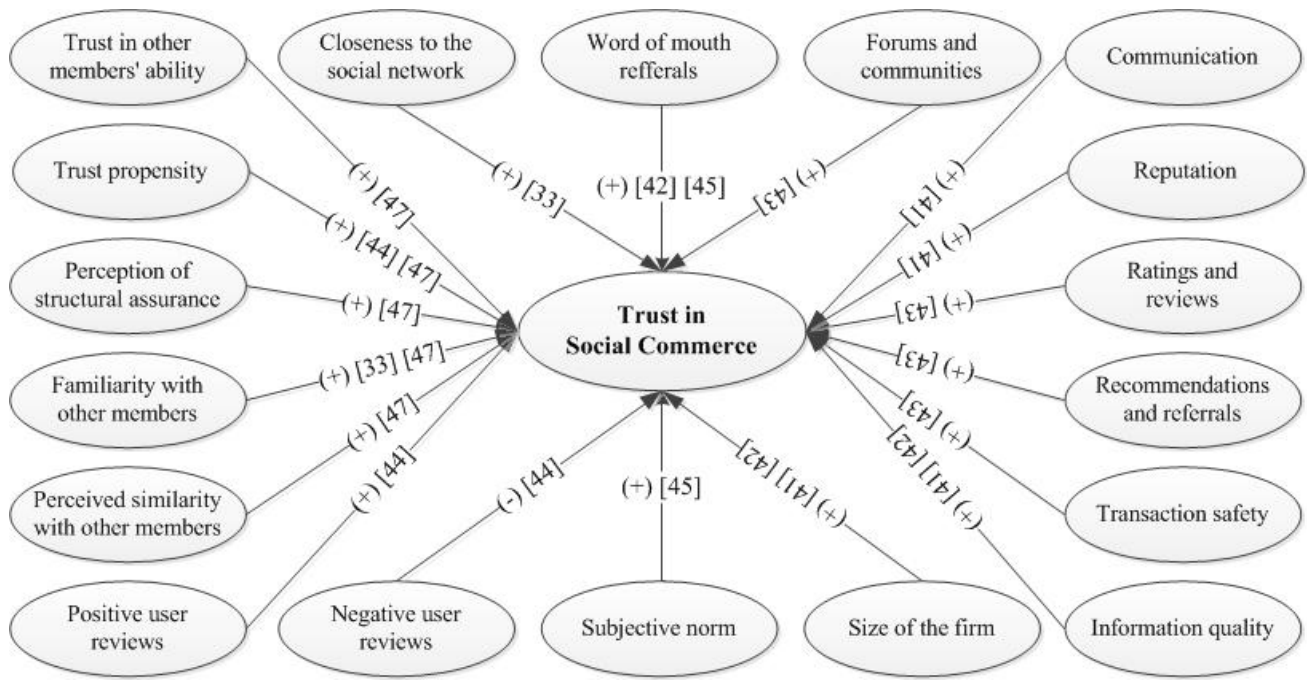

FIGURE I.        FACTORS AFFECTING TRUST IN S-COMMERCE DERIVED FROM THE PAPERS IN TABLE I